\documentclass[a4paper]{article}

\usepackage{INTERSPEECH2022}

\usepackage[abbreviations,nonumberlist,automake]{glossaries-extra}

\makeglossaries
\newacronym{srt}{SRT$_{50}$}{Speech Recognition Threshold for 50 \% correctly understood words}
\newacronym{sii}{SII}{Speech Intelligibility Index}
\newacronym{fade}{FADE}{Framework of Auditory Discrimination Experiments}
\newacronym{snr}{SNR}{Signal to Noise Ratio /dB}
\newacronym{rms}{RMS}{Root Mean Square error}
\newacronym{bif}{BIF}{Band Importance Function of the speech intelligibility index. Sometimes also called Frequency Importance Functions}
\newacronym{nns}{NNS}{band importance function various NoNSense syllable tests where most of the english phonemes occur equally often}
\newacronym{asr}{ASR}{Automatic Speech Recognizer}
\newacronym{hmmgmm}{HMM-GMM}{Hidden Marcov Model in conjunction with a Gaussian Mixture Model}
\newacronym{sgbfb}{SGBFB}{Separable Gabor FilterBank}
\newacronym{mfcc}{MFCC}{Mel Frequency Cepstral Coefficients}
\newacronym{icra5-250}{ICRA5-250}{modulated speech noise with a long term spectrum of a male speaker at normal vocal effort as defined by the International Collegium for Rehabilitation Audiology \cite{Dreschler2001} with a duration of pauses no longer than 250ms as proposed by \cite{Wagener2006}}
\newacronym{icra1}{ICRA1}{unmodulated speech noise with a long term spectrum of a male speaker at normal vocal effort as defined by the International Collegium for Rehabilitation Audiology \cite{Dreschler2001}}
\newacronym{stft}{STFT}{Short Time Fourier Transform}
\newacronym{yue}{YUE}{a dialect of Cantonese that is spoken in Hong Kong}

\newlength\maxlength
\newlength\thislength
\newglossarystyle{mystyle}
{%
  {
    \setlength{\maxlength}{0pt}%
    \forglsentries[\currentglossary]{\thislabel}%
    {%
       \settowidth{\thislength}{\glsentryshort{\thislabel}}%
       \ifdim\thislength>\maxlength
         \setlength{\maxlength}{\thislength}%
       \fi
    }%
    \settowidth{\thislength}{\hspace{1.5em}=\hspace{1em}}%
    \setlength{\glsdescwidth}{\linewidth-\maxlength-\thislength-2\tabcolsep}%
    \begin{tabular}{l@{\hspace{1.5em}\hspace{1em}}p{\glsdescwidth}}
    \toprule
    \multicolumn{1}{l}{\textbf{Abbreviation}} &
    \multicolumn{1}{@{}l}{\textbf{Meaning}}\\%
    \midrule
  }%
  {
     \bottomrule
     \end{tabular}%
  }%
  \renewcommand*{\glsgroupheading}[1]{}%
}

\usepackage{mathrsfs}

\usepackage[textwidth=15mm]{todonotes}

\newcommand{\col}{black}

\title{Lombard Effect for Bilingual Speakers in Cantonese and English:\\importance of spectro-temporal features}
\name{Maximilian Karl Scharf$^1$,Sabine Hochmuth$^2$, Lena L.N. Wong$^3$, Birger Kollmeier$^4$, Anna Warzybok$^5$}
\address{
  $^{1,2,4,5}$Medical physics and cluster of excellence Hearing4all, CvO university Oldenburg, Germany\\
  $^3$Clinical Hearing Sciences (CHearS) Laboratory, Faculty of Education, The University of Hong Kong, Hong Kong SAR, China}
\email{maximilian.scharf@uni-oldenburg.de, anna.warzybok-oetjen@uni-oldenburg.de}

\begin{document}

\maketitle
\begin{abstract}
 For a better understanding of the mechanisms underlying speech perception and the contribution of different signal features, computational models of speech recognition have a long tradition in hearing research. Due to the diverse range of situations in which speech needs to be recognized, these models need to be generalizable across many acoustic conditions, speakers, and languages. This contribution examines the importance of different features for speech recognition predictions of plain and Lombard speech for English in comparison to Cantonese in stationary and modulated noise. While Cantonese is a tonal language that encodes information in spectro-temporal features, the Lombard effect is known to be associated with spectral changes in the speech signal. These contrasting properties of tonal languages and the Lombard effect form an interesting basis for the assessment of speech recognition models. Here, an automatic speech recognition-based (\gls{asr}) model using spectral or spectro-temporal features is evaluated with empirical data. The results indicate that spectro-temporal features are crucial in order to predict the speaker-specific speech recognition threshold \gls{srt} in both Cantonese and English as well as to account for the improvement of speech recognition in modulated noise, while effects due to Lombard speech can already be predicted by spectral features. 
\end{abstract}


\section{Introduction}

\subsection{General aim}
\textcolor{\col}{
The aim of this study is to characterize the speech features involved in the Lombard effect for a comparison between a tonal language and a non-tonal language while minimizing effects due to variation between speakers.}

\subsection{Lombard Effect}
Discovered in 1911 \cite{Lombard1911}, the Lombard effect describes an increase in amplitude, duration as well as a change in the spectrum of speech, together with a change in visual cues when the speaker is introduced to loud sound fields. More precisely, the Lombard effect leads to an upward shift of the fundamental F0 frequency and formants F1, F2 as well as an increased average vowel duration \cite{Alghamdi2018}. This change in speech production is also referred to as increased vocal effort. Additionally, it has been reported by \cite{Junqua1993} that the spectral change of increased vocal effort leads to improved speech recognition. 

\subsection{Tonal and non-tonal languages}
Tonal languages, such as Cantonese, are spoken by a large fraction of the human world population and differ from non-tonal languages, such as English, in the property that information is encoded by changes in pitch within syllables. Thus, modeling a tonal encoding of information requires a model that incorporates both spectral as well as temporal features as an internal representation of speech signals. Established, and highly successful models for non-tonal languages are commonly used to describe speech recognition by spectral features only \cite{ANSI1997}. In the past, and with some success, these models have also been applied to tonal languages \cite{Chen2016}. 

\subsection{Spectral and spectro-temporal features of speech}
The current literature (see above) suggests that an improvement of the speech recognition threshold due to Lombard speech is connected to spectral changes of speech, while tonal languages make use of spectro-temporal coding of information which remains mostly unchanged during Lombard speech.

\textcolor{\col}{Acoustic speech features are commonly derived from a short time Fourier transform (\gls{stft}) with logarithmically scaled frequency and amplitudes. For the frequency, an approximation of the perceptually motivated mel scale is typically used, which differs from $\mathrm{log}_{10}$ in the base of the logarithm as well as in the vertical and horizontal intercept \cite{ETSI2003}:}

\begin{equation}
    \mathrm{Mel}(f)=\mathrm{log}_{10^{1/2595}}(1+\frac{f}{700})
\end{equation}

Automatic speech recognition systems can utilize this spectrum of a time-windowed section of a signal as feature vector, which is updated periodically for a fixed time shift and can be displayed on a 2D-grid as a spectrogram, called log-mel-spectrogram. A drawback of this data representation is that temporal changes do not enter the feature vector and need to be included by means of derivatives with respect to time.

One method to circumvent this problem is to calculate the Mel Frequency Cepstral Coefficients \gls{mfcc}. \gls{mfcc} take the log-mel-spectrogram and calculate the real part of its Fourier transform. \textcolor{\col}{Cepstra contain information about the temporal structure of a signal and allow to deconvolve, e.g., the glottal excitation signal from the vocal tract filter and the room impulse response  that are characteristic for different quefrencies. If more information about the temporal structure beyond the range of the time window of the \gls{stft} is required, derivatives of the \gls{mfcc} with respect to time are sometimes included in the feature vector.}







Another method of collecting temporal information of speech for a feature vector is the use of Separable GaborFilter Banks, short \gls{sgbfb} \cite{Schaedler2015a}. These work like a set of activation maps for the log-mel-spectrogram under the constraint that the resulting features are uncorrelated. \gls{sgbfb} can be calculated efficiently, because of their property that the required 2D-filter can be separated into a filter for the time- and frequency- axis. Due to the fact that these activation maps spread over many time samples, it is expected that \gls{sgbfb} contain more relevant temporal information for speech recognition than \gls{mfcc}. \gls{sgbfb} have been shown to be superior to \gls{mfcc} for accurate simulations of speech recognition in fluctuating noise conditions \cite{Schaedler2016c}.

\subsection{Influence of noise on speech recognition}
Besides the characteristics of the tested speech, the spectro-temporal structure of the masking noise is relevant for the resulting speech recognition threshold for which 50\% of words are understood correctly (\gls{srt}). Generally, continuous noise leads to higher \gls{srt} than noise with a spectro-temporal structure, i.e., fluctuating noise, where prospective listeners are able to make extensive use of information that is present in sections of low \gls{snr}, which is also referred to as ''listening in the gaps''. For the purpose of this study, two types of a standardized noise were used \cite{Dreschler2001,Wagener2006}: \gls{icra1}, which is a stationary signal with speech like spectrum, and  \gls{icra5-250}, which has a temporal structure with gaps no longer than 250ms, such that the \gls{srt} of icra5-250 is expected to be lower than for \gls{icra1}.

\subsection{Hypothesis}
This study aims to model the absolute \gls{srt} together with the Lombard Gain, which is the difference between \gls{srt} in Plain and Lombard speech, for bilingual speakers of Cantonese and English. Because of the different speech features that are relevant for the prediction of the Lombard gain and an absolute \gls{srt} prediction of tonal languages, the following hypotheses are tested:

\begin{enumerate}
    \item spectro-temporal features are important to describe the speaker specific \gls{srt} across different types of masker while
    \item spectral features suffice to describe the Lombard gain.
\end{enumerate}

\section{Methods}

\subsection{Empirical Data}
Two female and two male bilingual talkers recorded the matrix sentence test corpus \cite{Kollmeier2015} in both Cantonese (Hong Kong variety \gls{yue}) and English for plain and Lombard speech each. The Lombard effect was induced by \gls{icra1} noise with 80 dB SPL. For obtaining the \gls{srt}, these matrix sentence recognition tests were conducted with 15 normal hearing listeners, who were monolingual natives of the given language, in two different noise conditions: \gls{icra1} and \gls{icra5-250}. The adaptive procedure with word-scoring used for this test was according to \cite{Brand2002}.
The absolute \gls{srt} of a speaker was calculated as the mean across all listeners, while the Lombard gain was calculated as the mean across the individual Lombard gain for each listener. The measurement error of the mean $\sigma_\mu$ was estimated by:

\begin{equation}
    \sigma_\mu\approx\sigma/\sqrt N
\end{equation}

Where $\sigma$ is the standard deviation across N listeners.

\subsection{FADE Modeling approach}
\label{par:methods:fade}
The Framework of Auditory Discrimination Experiments' (\gls{fade}) approach to speech recognition threshold modeling as proposed by \cite{Schaedler2015} is based on an automatic speech recognizer (\gls{asr}), which acts as a model of a well-trained human listener. The \gls{fade} approach supports arbitrary types of features, based on the speech signals, to enter the \gls{asr} back end. For the simulation, the publicly available \gls{fade} distribution \cite{Schaedler2021b} was used with default settings.

The \gls{asr}-based \gls{srt} prediction consisted of the following steps: first, the audio of each sentence of the used test lists was mixed with random sections of the masking noise for N different \gls{snr}. Second, N Hidden Marcov Models in conjunction with a Gaussian Mixture Model (\gls{hmmgmm}) were trained on the features that were derived from the labeled signals for each \gls{snr}. The log-mel-spectrogram, from which the features were calculated, was based on a \gls{stft} with 25ms hamming window and a time shift of 10ms. The underlying Marcov model assumed 16 consecutive states per word and four states for silence, beginning or end of a sentence. Additionally, only transitions between words according to the matrix test syntax were possible. As a third step a new set of K matrix test audio signals, mixed with random sections of noise for a range of K different \gls{snr}, was used to test the performance of the N differently trained models. The resulting model performances as a function of N training- and K testing-\gls{snr} were used to calculate the \gls{srt} in a last step: the lowest value of the K testing \gls{snr} for a rate of 50\% correctly identified words is returned as \gls{srt}. As the simulation is done for discrete values of testing \gls{snr}, the final prediction of the \gls{srt} is based on an interpolation of the data. Uncertainties on the model prediction originate from the size of the data set used for testing.

The \gls{fade} simulations were done for every combination of the two types of input features for the \gls{asr} (\gls{mfcc}, \gls{sgbfb}), speakers (4), language (ENG, \gls{yue}), vocal efforts (plain, Lombard), and masking noises (\gls{icra1}, \gls{icra5-250}), resulting in 64 computations.

\subsection{Statistical measures}
The predictive power of the simulations were quantified by the Pearson correlation coefficient R, the root mean square error \gls{rms}, bias of a linear regression and the $\chi^2/\nu$ measure with $\nu$ degrees of freedom.  $\chi^2/\nu$ respects both measurement- $\sigma_\mathrm{emp}$ and simulation- uncertainty $\sigma_\mathrm{sim}$:

\begin{equation}
    \chi^2=\sum_n \frac{\left(\mathrm{sim}_n-\mathrm{emp}_n\right)^2}{\sigma_{\mathrm{emp},n}^2+\sigma_{\mathrm{sim},n}^2}
\end{equation}

where sim$_n$ denotes the result of a single simulation with index n and emp$_n$ stands for the corresponding empirical measurement. 
\section{Results}

\begin{figure*}[t]
    \centering
    \includegraphics[width=\linewidth]{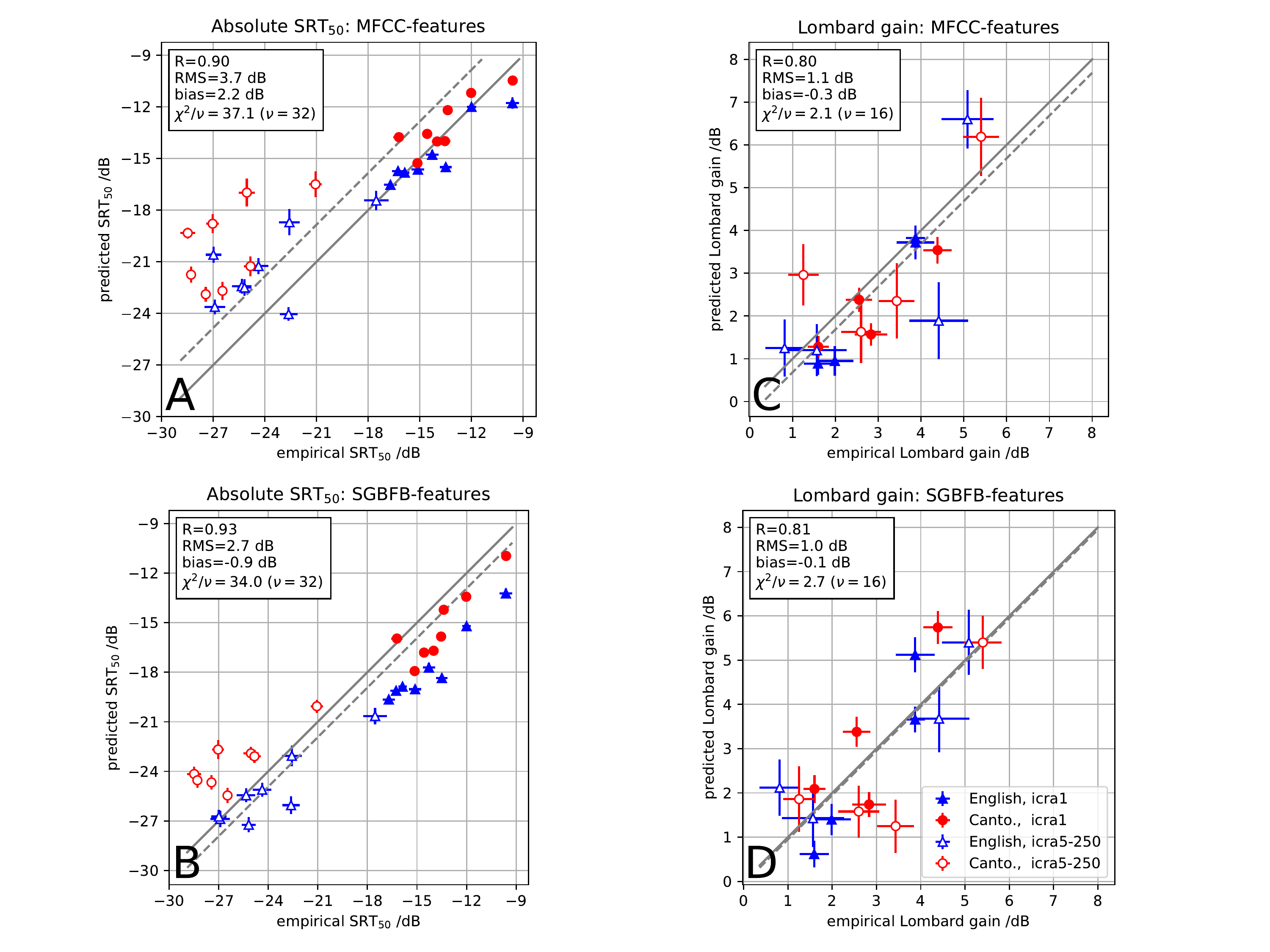}
    \caption{\glsfmttext{srt} and Lombard gain prediction of \gls{fade} with \glsfmttext{mfcc} (A and C, respectively) and \glsfmttext{sgbfb} features (B and D, respectively) \textcolor{\col}{for four bilingual speakers in stationary noise (\glsfmttext{icra1}) and fluctuating noise (\glsfmttext{icra5-250}). The solid line corresponds to perfect correlation, the dashed line depicts a linear interpolation.}}
    \label{fig:results}
\end{figure*}

\subsection{Simulation results}
The scatter plots of predicted vs empirically obtained speech recognition thresholds (\gls{srt}) are shown in figure \ref{fig:results}, where the prediction of the absolute \gls{srt} as well as Lombard Gain is displayed for \gls{mfcc}- and \gls{sgbfb}-based predictions. The predictions and empirical findings of absolute \gls{srt} correlate with R$\geq0.9$ for both feature sets.

Panel A and B show that the \gls{fade} prediction based on \gls{sgbfb} features was able to describe the data with smallest \gls{rms}, offset and $\chi^2/\nu$ where $\nu$ is the degrees of freedom of the modelled data. As the \gls{fade} prediction method is only working with the sound files, no information about the empirical \gls{srt} has flown into the prediction, such that $\nu$ is equal to the number of data points. 

\textcolor{\col}{The \gls{mfcc} based prediction in panel A showed a good agreement between simulations and empirical data in stationary noise \gls{icra1} (R=0.85). However, the correlation is weak for the \gls{icra5-250} noise condition (R=0.44), which is line with previous findings \cite{Schaedler2016c} that FADE can predict the \gls{srt} in fluctuating noise better with spectro-temporal feature sets (e.g., \gls{sgbfb}) than with either temporal or spectral feature sets in isolation. The \gls{mfcc} are not sufficient to take advantage of the temporal gaps in the masker as efficient as human listeners did. This observation holds for both languages. On the other hand, the predictions based on \gls{sgbfb} features of panel B coincide very well with the empirical data for both noise conditions. A negative bias of -2.6 dB for \gls{icra1} remained for \gls{sgbfb} features.}

Panel C and D show the Lombard gain, which is the difference between the \gls{srt} in plain and Lombard speech. Accordingly the remaining degrees of freedom of the data $\nu$ are cut in half. Both feature sets show high correlation with vanishing bias together with a value of $\chi^2/\nu$ close to unity, which means that the deviation of the model from perfect correlation with the measurement is in agreement with the uncertainties. It can be observed that the Lombard gain prediction was independent of the used features.

%
\section{Discussion}
The discussion will distinguish between the absolute \gls{srt} prediction and prediction of the Lombard gain according to the two initial hypotheses that spectro-temporal features are important to describe the speaker specific \gls{srt} while spectral features suffice to describe the Lombard gain.

\subsection{Absolute \glsfmttext{srt} prediction}

Panel A and B of figure \ref{fig:results} show the absolute \gls{srt} prediction for the two different speech features \gls{mfcc} in A and \gls{sgbfb} in B.

Based on the calculated statistical measures, the \gls{sgbfb} based prediction was superior to the \gls{mfcc}. Especially in the fluctuating noise condition \gls{icra5-250}, \gls{mfcc} based predictions failed to describe the empirical data, which is in agreement with known properties of \gls{mfcc} \cite{Schaedler2016c}. However, \gls{srt} in \gls{icra1} was correctly predicted with \gls{mfcc} for both languages, which means that the \gls{mfcc} features provided to the \gls{asr} backend of \gls{fade} were sufficient to predict \gls{srt} for both English and the tonal language Cantonese in stationary \gls{icra1} noise. On the other hand, the \gls{sgbfb} based prediction worked for all noise conditions, such that the individual \gls{srt} of a speaker was predicted with higher accuracy, which is in line with hypothesis 1, that spectro-temporal features are important to describe the individual \gls{srt}. Even so, a bias of -2.6 dB is present for the \gls{sgbfb} based prediction in the stationary \gls{icra1} noise, that is expected: \textcolor{\col}{\gls{fade} works as a model of an ideally trained listener, which may lead to a negative offset, if the listeners employed are neither highly trained nor selected to be free from any minor sign of a hearing deficiency. Note that according to \cite{Huelsmeier2021}, even normal-hearing listeners have to be modeled with a small processing deficit (denoted as level uncertainty) to obtain precise model predictions with FADE.}

Even though \gls{sgbfb} features resulted in the best prediction of the absolute \gls{srt}, as can be seen in figure \ref{fig:results} panel B, $\chi^2/\nu$ is not in the range of unity, which shows that the model, or the applied features, still need improvement to describe the data to full agreement with measurement uncertainties.

\subsection{Lombard gain prediction }

Panel C and D of figure \ref{fig:results} show the Lombard gain prediction for the two different speech features \gls{mfcc} in C and \gls{sgbfb} in D. 

As can be seen from the statistical measures with similar correlation coefficient, \gls{rms} of about 1 dB and bias below 0.3 dB, the two prediction methods are equivalent in their predictive power. $\chi^2/\nu$ is close to unity for both models, which means that the deviation of the prediction from perfect correlation with the empirical data is in agreement with measurement uncertainties together with the precision of the model. It is evident that the Lombard gain, which is associated with spectral changes of speech, can be adequately modelled by \gls{mfcc}. \gls{mfcc} can only contain information from within the time window of the \gls{stft}, which was in this case 25ms. These findings provide evidence for hypothesis 2 that spectral features suffice to describe the Lombard gain.


\section{Conclusions}
The present study has shown that \gls{asr} based models of speech recognition such as \gls{fade} are a powerful tool for the prediction of \gls{srt} in stationary and fluctuating noise conditions for Cantonese and English. However the choice of input features is important for a correct prediction of the individual \gls{srt} of a speaker. \gls{sgbfb} features, which are similar to activation maps for the log-mel-spectrogram, were superior to \gls{mfcc}, which can only contain information limited by the time window of the \gls{stft}. On the other hand, it was shown that the Lombard gain can be predicted by both \gls{mfcc} as well as \gls{sgbfb}, because changes in \gls{srt} due to the Lombard effect appear to be mainly associated to a spectral change of speech. In summary it was found that:
\begin{enumerate}
    \item spectro-temporal features such as \gls{sgbfb} are important to predict the \gls{srt} for both Cantonese and English for speakers in stationary and in modulated noise.
    \item spectral features such as \gls{mfcc} suffice to describe the Lombard gain of a speaker, for both stationary and modulated noise.
\end{enumerate}
Although \gls{sgbfb} have been shown to be superior to \gls{mfcc} for individual \gls{srt} prediction, the model was not able to compute predictions with maximum possible agreement with empirical data, which will require further investigations of the model or the used features.

Furthermore, the aspect of differences between tonal and non-tonal languages has not yet been fully explored in this contribution. Since bilingual speakers were used, further analysis of the data may uncover if significant differences in the Lombard Gain for Cantonese and English exist even for the same speaker. This might be related to the usage of different speech cues for enhancing speech recognition via the Lombard effect across languages.
 
\section{Abbreviations}
\printabbreviations[title=,style=mystyle] 
\section{Funding}
This work was supported by the DFG grant ''Experiments and models of speech recognition across tonal and non-tonal language systems (EMSATON)'' grant number 415895050.
%

\section{Declaration of Conflicting Interests}
The Authors declare that there is no conflict of interest.

\clearpage
\bibliographystyle{IEEEtran}
\bibliography{Bibliography}

\end{document}